\documentclass{optica-article_test}

\journal{oe}

\articletype{Research Article}
\usepackage{caption}
\usepackage{subcaption}
\usepackage{lineno}
\usepackage{amsmath}
\usepackage{textgreek}
\usepackage{graphicx}
\usepackage{booktabs}
\usepackage{hyperref}
\hypersetup{
    colorlinks=true,
    linkcolor=blue,
    filecolor=magenta,      
    urlcolor=blue,
    }
    
\begin{document}

\title{Injection-seeded high-power Yb:YAG thin-disk laser stabilized by the Pound-Drever-Hall method}

\author{Manuel Zeyen,\authormark{1,*} Lukas Affolter,\authormark{1} Marwan Abdou Ahmed,\authormark{2} Thomas Graf,\authormark{2} Oguzhan Kara,\authormark{1} Klaus Kirch,\authormark{1,3} Adrian Langenbach,\authormark{1} Miroslaw Marszalek,\authormark{1} François Nez,\authormark{4} Ahmed Ouf,\authormark{5} Randolf Pohl,\authormark{5,6} Siddharth Rajamohanan,\authormark{5} Pauline Yzombard,\authormark{4} Aldo Antognini,\authormark{1,3,*} and Karsten Schuhmann\authormark{1}}

\address{\authormark{1}Institute for Particle Physics and Astrophysics, ETH, 8093 Zurich, Switzerland.\\
\authormark{2}Institut für Strahlwerkzeuge, Universität Stuttgart, Pfaffenwaldring 43, 70569 Stuttgart, Deutschland.\\
\authormark{3}Laboratory for Particle Physics, Paul Scherrer Institute, 5232 Villigen, Switzerland.\\
\authormark{4}Laboratoire Kastler Brossel, Sorbonne Université, CNRS, ENS-Université PSL, Collège de France, 75252 Paris Cedex 05, France.\\
\authormark{5}QUANTUM, Institute of Physics, Johannes Gutenberg-Universität Mainz, 55099 Mainz, Germany.
\authormark{6}Excellence Cluster PRISMA+, Johannes Gutenberg-Universität Mainz, 55099 Mainz, Germany.\\
}

\email{\authormark{*}zeyenm@phys.ethz.ch} 
\email{\authormark{*}aldo.antognini@psi.ch}

\section*{Abstract:}
We demonstrate an injection-seeded thin-disk Yb:YAG laser at 1030 nm, stabilized by the Pound-Drever-Hall (PDH) method. We modified the PDH scheme to obtain an error signal free from Trojan locking points, which allowed robust re-locking of the laser and reliable long-term operation. The single-frequency pulses have 50~mJ energy (limited to avoid laser-induced damage) with a beam quality of M²~<~1.1 and an adjustable length of 55-110 ns. Heterodyne measurements confirmed a spectral linewidth of 3.7~MHz. The short pulse build-up time (850~ns) makes this laser suitable for laser spectroscopy of muonic hydrogen, pursued by the CREMA collaboration.

\section{Introduction}

Thin-disk lasers (TDL) are well known for their high power and energy output at very good beam quality. High average power systems with close to diffraction limited beam quality and pulse durations from nanoseconds to femtoseconds are commercially available and a large variety of high-energy and high-intensity systems have been demonstrated \cite{negel_ultrafast_2015, Kuhn_2015,saltarelli_350-w_2019, kretschmar_thin-disk_2020, Wandt:20,radmard2022400,Dai2022}. While high-energy single transverse mode operation in TDLs is well established, high-energy single axial mode operation (i.e. single-frequency operation) is rarely pursued. Some projects using Yb:YAG are worth noting \cite{hankla_q-switched_2002, larionov_single-frequency_2005, paa_fast_2011, schuhmann_vbg_2013, stolzenburg_power_2005} but the vibrations caused by the impingement cooling of the disk make stable single-frequency operation in these high-power systems challenging.

In this paper we present an injection-seeded, Q-switched 1030~nm Yb:YAG TDL. Prior to trigger, a narrow-bandwidth CW seed laser is resonantly coupled into the TDL resonator, so that only a single axial mode is initially populated by the seed laser. The resulting high-energy pulse building up in the TDL is thus also narrow-bandwidth ("single-frequency"). Resonant injection of the seed light was achieved by stabilizing the TDL resonator length with the Pound-Drever-Hall (PDH) method \cite{pound_electronic_1946, drever_laser_1983}, and we achieved close to time-bandwidth limited pulses with energies over 50~mJ in $\text{TEM}_{00}$ mode, 55~ns pulse duration and < 5~MHz bandwidth. We reached pulse repetition rates up to 1~kHz at 14~mJ pulse energy. To the best of our knowledge, this is the first time a high-power TDL has been frequency-stabilized via the PDH-method, and the obtained single-frequency pulses have the highest energy to date.

In addition to delivering narrow bandwidth pulses, the here presented laser must fulfill a number of additional requirements, as it is part of the laser system used for the HyperMu experiment \cite{amaro_laser_2022, nuber2022diffusion} currently pursued at the Paul Scherrer Institut in Villigen, Switzerland. In this experiment, the CREMA collaboration \cite{pohl_size_2010, antognini_proton_2013, pohl2016laser, krauth_measuring_2021, thecremacollaboration2023helion} aims at the first laser spectroscopy measurement of the ground-state hyperfine splitting in muonic hydrogen. The laser system is based on a TDL delivering high-energy pulses at 1030~nm which are down-converted in a series of optical parametric oscillators (OPO) and optical parametric amplifiers (OPA) to the mid-IR region (around 6.8~\textmu m) where the hyperfine transition (100~MHz linewidth at full width half maximum) is expected. For efficient and reliable multi-stage down-conversion, the TDL output beam must be diffraction limited, must have stable beam parameters (pointing and beam size), and must have small pulse-to-pule energy variation. Additionally, for the spectroscopy experiment, the bandwidth must be $\leq$\,10~MHz, calling for pulse lengths on the order of 50~ns, and the laser must deliver the pulse with a short latency after trigger (about 1~\textmu s) owing to the muon lifetime of 2~\textmu s.

The TDL technology was chosen mainly since it allows short pulse build-up times, large pulse energies and good beam quality, as shown in earlier work of our group \cite{antognini_thin-disk_2009,Schuhmann:13,schuhmann_thin-disk_2017,zeyenthin}. The design of our new TDL presented here is based on three key innovations: 1) A PDH-based injection-seeding process, modified so that the TDL resonator is locking correctly on resonance with the seed even after large perturbations \cite{zeyen2023pound}. This makes the injection-seeding process very robust. 2) Asymmetric closing and opening of the resonator to obtain Gaussian pulses. For this purpose the outcoupling of the TDL resonator is voltage controlled via a Pockels cell, of which we control both electrodes independently. 3) A feedback loop controlling the beam pointing to mitigate slow drifts, and a feed-forward system stabilizing the pulse energy on a pulse-to-pulse level by dynamically adjusting the pulse build-up time.

The paper is organized as follows: Section~\ref{sec:ResonatorDesign} describes the TDL resonator design. Section~\ref{sec:Setup} describes the injection-seeding process and the Q-switching dynamics. Single-frequency operation is demonstrated in Section~\ref{sec:SingleFreq}, and the overall laser performance is presented in Section~\ref{sec:LaserPerformance}.

\section{Resonator design \label{sec:ResonatorDesign}}

The design of our TDL is based on the TDL that was employed in the laser spectroscopy of the muonic helium ion \cite{schuhmann_thin-disk_2017, krauth_measuring_2021}. One of the main design criteria was to ensure output beam parameters insensitive to variations of the thermal lens of the thin-disk. Such stability can be achieved by setting up the resonator as a Fourier transform (FT) resonator, in which the mode undergoes an optical FT in one round-trip from disk to disk. Such a FT resonator has the round-trip ABCD matrix \cite{schuhmann_multipass_2018}
\begin{equation}
M_{\textnormal{FT}}=\left(\begin{array}{cc}
0 & F\\
-\frac{1}{F} & 0
\end{array}\right),\label{eq:M_FT}
\end{equation}
where $F>0$ is the so-called Fourier parameter of the resonator. The resulting beam size of the $\textnormal{TEM}_{00}$ eigenmode at the disk is given by \cite{schuhmann_multipass_2018}
\begin{equation}
w_0 = \sqrt{\frac{F\lambda}{\pi}},
\end{equation} with $\lambda$ being the laser wavelength (in our case $\lambda = 1030$~nm). For example, a resonator sustaining a mode with $w_0 = 2.1$~mm has $F=13.4$~m (it turns out that $F$ is precisely the Rayleigh length of the FT resonator eigenmode at the disk).
\begin{figure}
\centering
\includegraphics[width=0.7\textwidth]{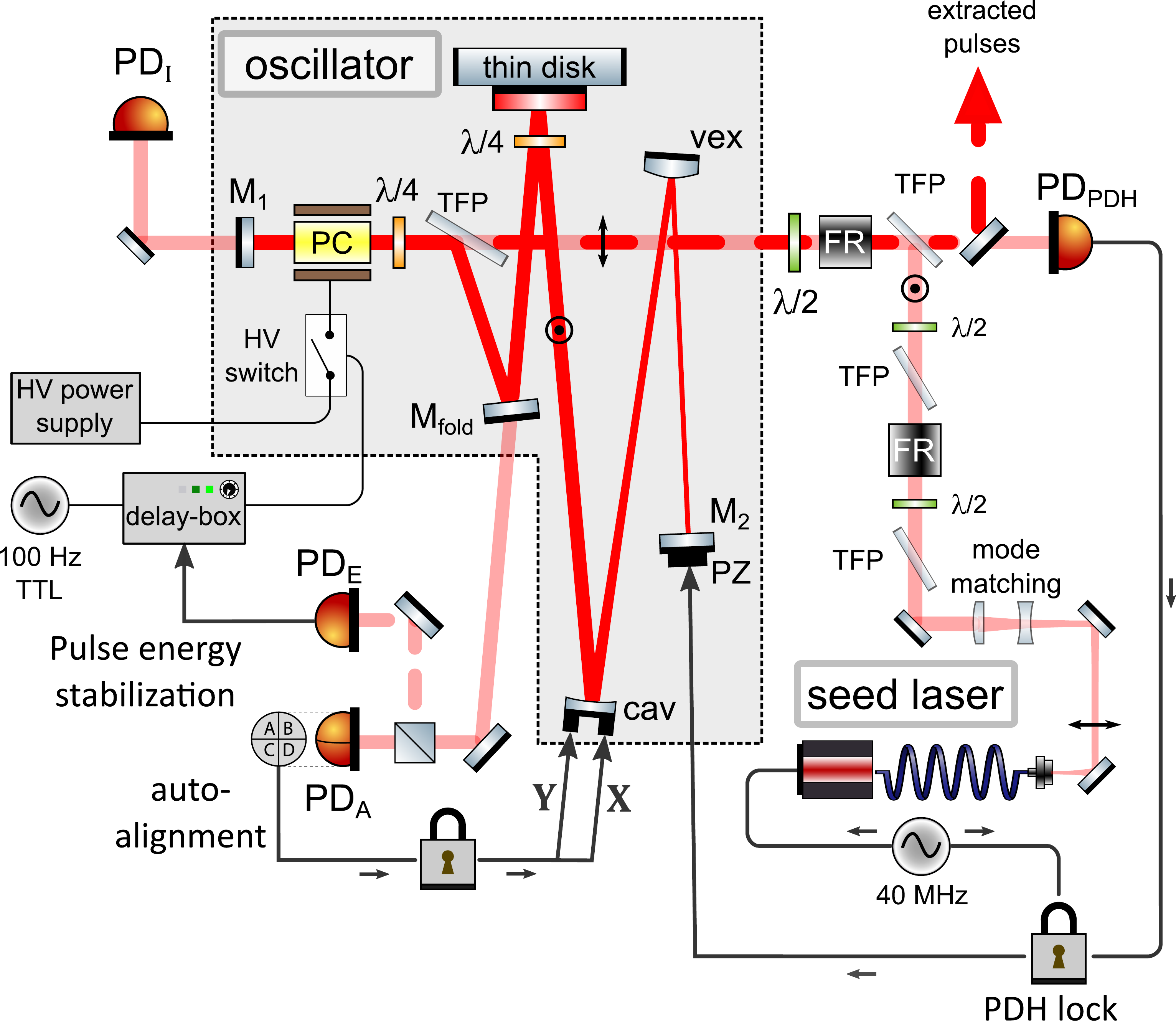}
\caption{Schematic of the injection-seeded thin-disk oscillator. HV: high-voltage, PC: Pockels cell, TFP: thin-film polarizer, FR: Faraday rotator, PD: photodiode, PDH lock: Pound-Drever-Hall locking scheme, $\updownarrow$: P-polarization, $\odot$: S-polarization, \textlambda /2: half-wave plate, \textlambda /4: quarter-wave plate. The resonator is spanning between end-mirrors $\text{M}_1$ and $\text{M}_2$. The PC-\textlambda /4-TFP system forms a HV controlled outcoupler steering the Q-switching dynamics of the oscillator. Through this TFP the seed laser output is injected into the resonator prior to trigger, and the pulses are extracted from the oscillator after pulse-buildup. Resonant coupling of the seed laser light into the oscillator is achieved through a PDH-locking scheme acting on the resonator length. The resonator alignment is stabilized via an autoalignment system including the quadrature photodiode $\text{PD}_\text{A}$ and two piezos steering the alignment of mirror "cav". The pulse energy is stabilized via a feed-forward system including the photodiode $\text{PD}_\text{E}$ and a delay box adjusting the pulse build-up time.}
\label{fig:OsciSketch}
\end{figure}

We realized the FT resonator as a telescopic resonator in a V shape with the disk at the tip of the V as shown in Fig.~\ref{fig:OsciSketch}. This configuration reduces the pulse build-up time for two reasons: the Galilean telescope shortens the physical propagation length of the resonator while the V-shape doubles the number of passes over the disk per round-trip. The disk (thickness: 275~\textmu m, Yb-doping: 7~at.\,\%) is continuously pumped at 969~nm (zero-phonon line) with about 350~W and water-cooled from the back side. The pump spot diameter is set to $d_\text{P} = 5.8$~mm to ensure efficient energy transfer from the disk to the laser mode. A quarter-wave plate is placed in front of the disk to suppress spatial hole-burning as detailed in \cite{schuhmann2018spatial}.

\begin{figure}
	\begin{minipage}{0.35\textwidth}
		\centering
		\begin{tabular}{lr}
			\toprule
			Parameter        & value\\
			\midrule
			$f_\text{D}$ & $\approx 10$~m\\
			$f_\text{cav}$ & 0.5~m \\
			$f_\text{vex}$ & -0.15~m \\
			\midrule
			$d_0$ & 0.55~m \\
			$d_1$ & 0.5~m \\
			$d_2$ & 0.34~m \\
			$d_3$ & 0.45~m \\
			$d_{\text{tot}}$ & 1.89~m \\
			\midrule
			$w_0$ & 2.1~mm \\
			$d_\text{P}$ & 5.8~mm \\
			\bottomrule
		\end{tabular}
		\captionof{table}{Resonator parameters.}
		\label{tab:params}
	\end{minipage}
	\hfill
	\captionsetup{width=0.6\textwidth}	
	\begin{minipage}{0.63\textwidth}
		\centering
		\includegraphics[width=0.97\linewidth]{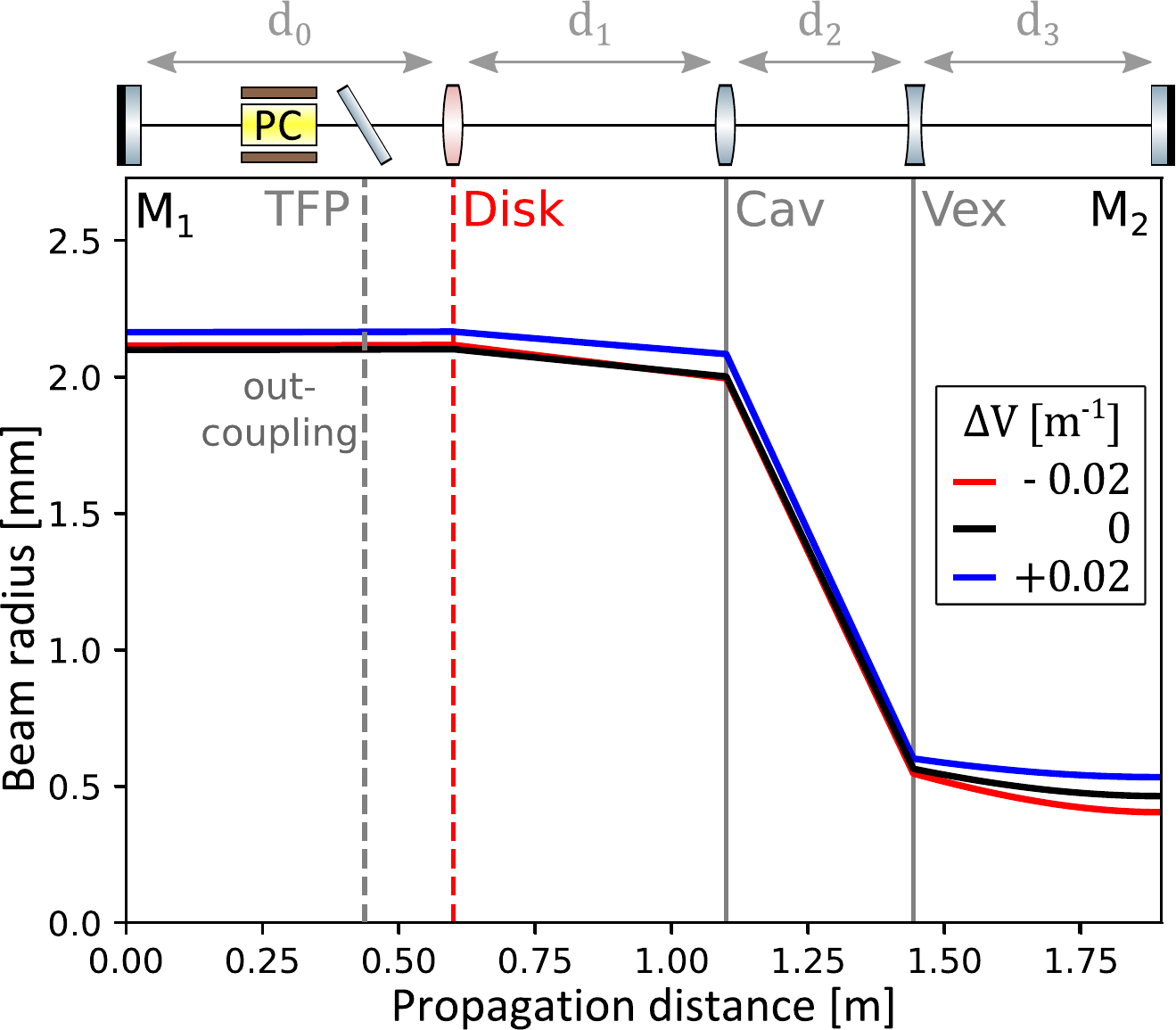}
		\captionof{figure}{Beam radius (1/$\text{e}^2$) of the $\text{TEM}_{00}$ mode along the TDL resonator for variations of the disk dioptric power $\Delta V$. For simplicity the location of folding mirror $\text{M}_\text{fold}$ is not shown.}
		\label{fig:beamsize}
	\end{minipage}
\end{figure}

In contrast to our earlier FT resonator \cite{antognini_thin-disk_2009, schuhmann_thin-disk_2017}, we used a quasi flat disk with focal length $f_\text{D} \approx 10$~m instead of a curved disk (which is often employed in telescopic thin-disk lasers). The disk is thus not part of the telescope, and an additional concave mirror must be introduced. A round-trip in the resonator follows the propagation (see Fig.~\ref{fig:OsciSketch} and Fig.~\ref{fig:beamsize})
\begin{equation}
\text{M}_{1}~\overset{d_{0}}{\text{ \textemdash }}\text{ disk }\overset{d_{1}}{\text{ \textemdash }}\text{ cav }\overset{d_{2}}{\text{ \textemdash }}\text{ vex }\overset{d_{3}}{\text{ \textemdash }}~\text{M}_{2}~\overset{d_{3}}{\text{ \textemdash }}\text{ vex }\overset{d_{2}}{\text{ \textemdash }}\text{ cav }\overset{d_{1}}{\text{ \textemdash }}\text{ disk}\overset{d_{0}}{\text{ \textemdash }}~\text{M}_{1},
\end{equation}
where cav and vex stand for the concave and a convex mirror with focal lengths $f_\text{cav}$ and $f_\text{vex}$, respectively. The flat end-mirrors are denoted $\text{M}_{1}$ and $\text{M}_{2}$, and the mirror spacings are defined as $d_0$, $d_1$, $d_2$ and $d_3$.

Since the beam is large and collimated between end-mirror $\text{M}_1$ and the disk, $d_0$ can be chosen arbitrarily (as long as $d_0$ is much smaller than the Rayleigh length of the eigenmode in this part of the resonator) and thus does not enter the calculations. The distance $d_1$ between disk and concave mirror was chosen as a compromise between short propagation and small incidence angle on the concave mirror to mitigate astigmatism. Finally, the remaining degrees of freedom $d_2$ and $d_3$ are determined by matching the ABCD round-trip matrix of the resonator to the FT matrix \eqref{eq:M_FT}. Solving the equations leads to

\begin{equation}
d_2 = f_\text{vex}
+
\frac{F \left( f_\text{D} - d_1 \right) }{ F - f_\text{D} - \xi \left( F -  f_\text{D}\right) + F f_\text{D}/f_\text{cav}}
+
d_1 \frac{f_{\text{D}}}{F -f_{\text{D}} - \xi \left( F - f_{\text{D}} \right) + F f_{\text{D}}/f_{\text{cav}}},
\label{eq:d2_xi}
\end{equation}

\begin{equation}
d_3 = f_{\text{vex}} 
+
\frac{F f_\text{vex}^2}{2} \left[ \frac{1}{f_{\text{D}^2}} - \frac{1}{F^2}
+  \frac{1}{f_{\text{cav}}^2} \right] \left( 1-\xi \right)^2
+
F \left( \frac{f_{\text{vex}}}{f_\text{cav}}\right)^2 \left[ \left( \xi + \frac{f_{\text{cav}}}{f_{\text{D}}} \right) \left( 1-\xi \right) + \frac{1}{2} \xi^2 \right],
\label{eq:d3_xi}
\end{equation}
where we defined the parameter $\xi = d_1/f_{\text{cav}}$.

If a sufficiently curved disk is used, the concave mirror is not required so that $f_\text{cav}\rightarrow \infty$ and $d_1\rightarrow 0$, with $\xi \rightarrow 0$. Thus, Eq.~\eqref{eq:d2_xi} \& Eq.~\eqref{eq:d3_xi} simplify to
\begin{equation}
d_2 \rightarrow f_\text{vex} + \frac{f_\text{D} F}{F-f_\text{D}}
\end{equation}
and
\begin{equation}
d_3 \rightarrow f_\text{vex} + \frac{F}{2}\left[ \left(\frac{f_\text{vex}}{f_{\text{D}}}\right)^2 - \left( \frac{f_{\text{vex}}}{F} \right) ^2 \right].
\end{equation}
The chosen values for all focal lengths and propagation distances are given in Table \ref{tab:params}. Figure \ref{fig:beamsize} illustrates the simulated beam size in the TDL resonator for the nominal focal power of the disk (black solid line). It also depicts the sensitivity of the mode in case of thermal lensing where the disk picks up an additional focal power of $\Delta V = \pm0.02 \ \text{m}^{-1}$ (blue and red curves, respectively).

Both injection of the seed laser and pulse extraction are occuring through the thin-film polarizer (TFP) located between end-mirror $M_1$ and the disk where the influence of a varying thermal lens on the resonator mode is minimal. Indeed, in this region of the resonator, small variations $\Delta V$ lead to a relative change in mode size \cite{schuhmann_multipass_2018} of
\begin{equation}
\frac{w}{w_0} = 1 + \frac{1}{4} F^2 \Delta V^2 + \frac{5}{32}F^4 \Delta V^4 + \ldots,
\end{equation}
where $w_0$ is the mode size for the nominal value of $f_\text{D}=f_\text{nom}$ and $w$ is the mode size for $1/f_\text{D} = 1/f_\text{nom} \pm \Delta V$. The quadratic dependence of $w$ on $\Delta V$ results in stability against small variation of $\Delta V$ for both the pulse extraction and the seed incoupling.

An additional advantage of incoupling the seed through the TFP is that the reflectivity of the TDL resonator can be adjusted to optimize the PDH error signal. This greatly simplifies the commissioning of the laser and optimization of the PDH locking-system. 

The seed is protected from the intense laser pulses by two optical isolators consisting of TFPs and Faraday rotators (FR) as well as an internal isolator in the seed laser module. Q-switching of the TDL is performed by varying the resonator losses with the standard combination of a TFP, a quarter-wave plate and a Pockels cell (PC), denoted as TFP-$\lambda/4$-PC (see Subsec. \ref{subsec:QSwitching}).

\section{Experimental setup\label{sec:Setup}}
\subsection{Pound-Drever-Hall stabilized resonator length \label{subsec:PDH-lock}}

\begin{figure}
\centering
\includegraphics[width=0.6\textwidth]{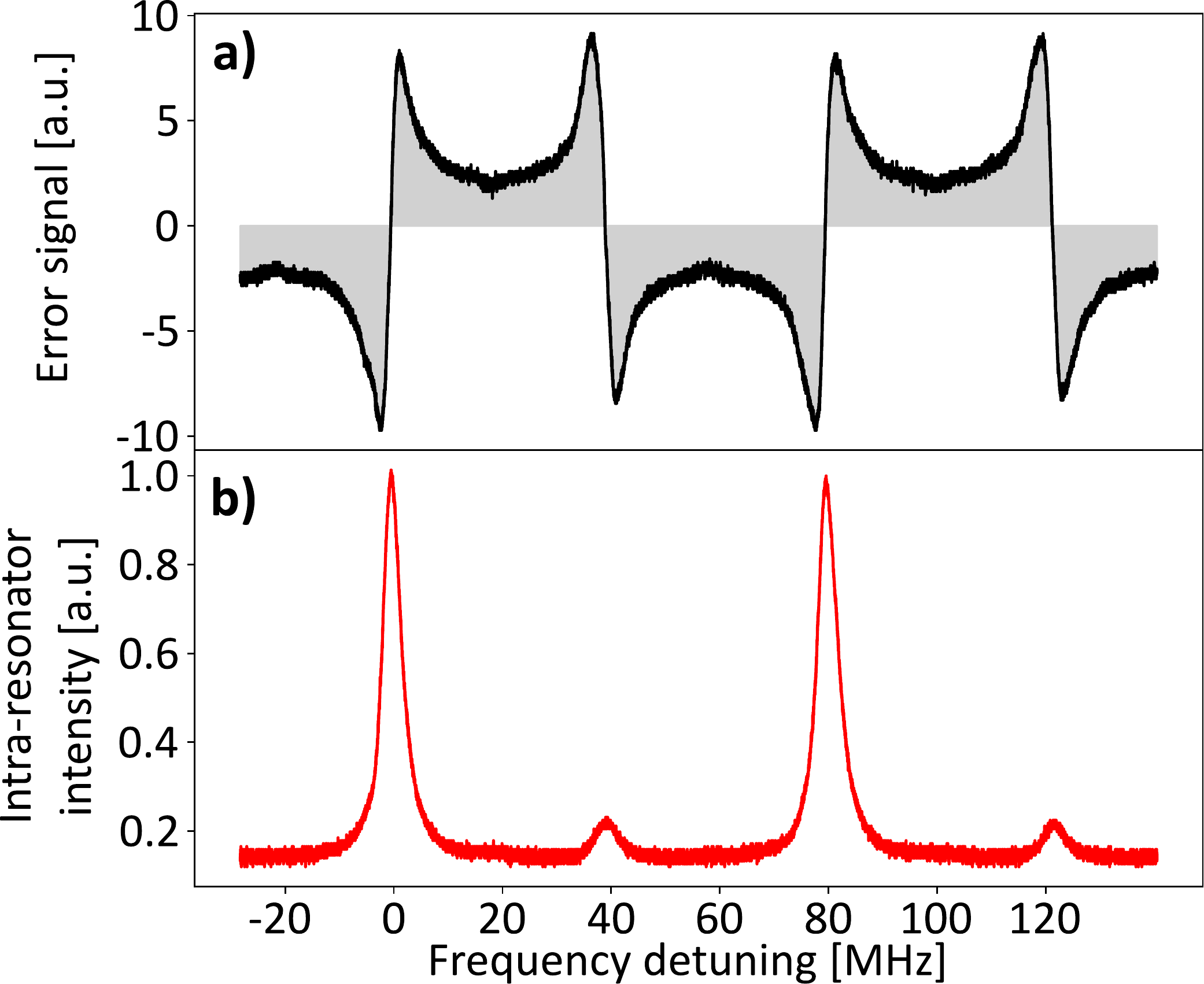}
\caption{\textbf{a)} Typical open-loop PDH-error-signal measured with photodiode $\text{PD}_\text{PDH}$ in Fig.~\ref{fig:OsciSketch} obtained by scanning the resonator length. The modulation frequency of the
seed laser is 40~MHz, i.e. half the free spectral range of the resonator. \textbf{b)} Corresponding $\text{TEM}_{00}$ intra-resonator intensity measured with $\text{PD}_\text{I}$. As the resonator length is scanned, the side bands are also visible at $\pm 40$~MHz.}
\label{fig:Our_PDH}
\end{figure}

Prior to a laser trigger, the TDL resonator is populated with seed laser photons (i.e. injection-seeded) by stabilizing the resonator length to the seed laser (Toptica DLPro at 1030~nm) frequency via PDH locking. We use a modified PDH-scheme that guarantees automatic locking on resonance even after large perturbations \cite{zeyen2023pound}. This is achieved by modulating the phase of the seed laser light at half the free spectral range (FSR) of the TDL resonator to obtain the error signal shown in Fig.~\ref{fig:Our_PDH}. This error signal has only one zero crossing between two neighboring resonances and is thus free from "Trojan" lock points \cite{zeyen2023pound}. This allows the servo-controller to always re-lock the TDL resonator correctly on resonance. Since the frequency of the intra-resonator light is defined by the seed laser frequency, it does not matter on which resonance the TDL is stabilized.

The seed laser is sinusoidally current-modulated at $f_\text{M} = 40$~MHz by using one channel of a function generator (Tektronix AFG1062). The second channel, a copy of the modulation signal, is used as reference (local oscillator) for demodulation in the PDH scheme. The phase delay between modulation signal and reference can be continuously tuned by adjusting the relative phase between the two channels directly on the function generator.

The reflected light used for the PDH-lock is detected as leakage light through a 45° high-reflective mirror by the fast photodiode denoted $\text{PD}_\text{PDH}$ in Fig.~\ref{fig:OsciSketch}. During pulse extraction, $\text{PD}_\text{PDH}$ saturates so that the lock is not functional for ca. $40$~\textmu s after pulse extraction~\cite{zeyen2023pound}. Once the photodiode has recovered, the resonator always re-locks correctly on resonance to the seed frequency thanks to our modified PDH scheme and is immediately ready for the next trigger. 

\subsection{Q-Switching dynamics \label{subsec:QSwitching}}
The time-dependent effective (power) reflectivity $R_\text{out}(t)$ of the system TFP-$\lambda/4$-PC can be calculated via the Jones vector formalism as

\begin{eqnarray}
R_\text{out}(t) & = & \left| \cos\left( 2\pi \frac{\Delta U(t)}{U_{\lambda/4}}\right)\cos\left( 2\alpha \right) 
-\frac{1}{2}\sin\left( 2\pi \frac{\Delta U(t)}{U_{\lambda/4}}\right)\sin\left( 4\alpha \right) \right.\\
 & & \left. + i \cdot \sin\left( 2\pi \frac{\Delta U(t)}{U_{\lambda/4}}\right)\sin\left( 2\alpha \right) 
\right|^2 \nonumber
\end{eqnarray}

where $\alpha$ is the rotation angle of the quarter-wave plate, $\Delta U(t) = U_1(t) - U_2(t)$ is the voltage difference between the two electrodes holding the BBO crystal of the PC in place, and $U_{\lambda/4}$ is the quarter-wave voltage of the PC.

\begin{figure}
\centering
\begin{subfigure}[b]{0.45\textwidth}
\centering
\includegraphics[width=\textwidth]{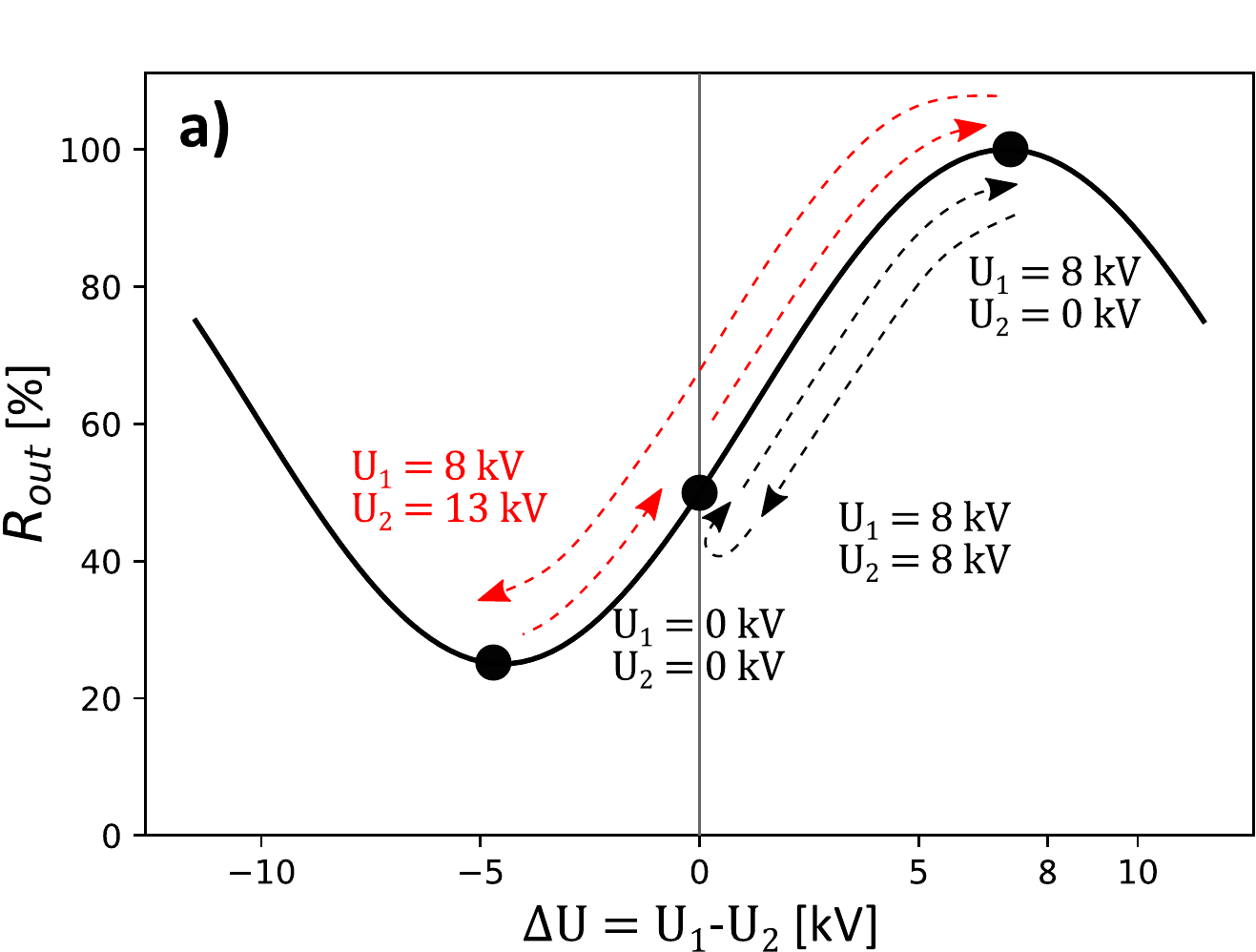}
\end{subfigure}
\hfill
\begin{subfigure}[b]{0.54\textwidth}
\centering
\includegraphics[width=\textwidth]{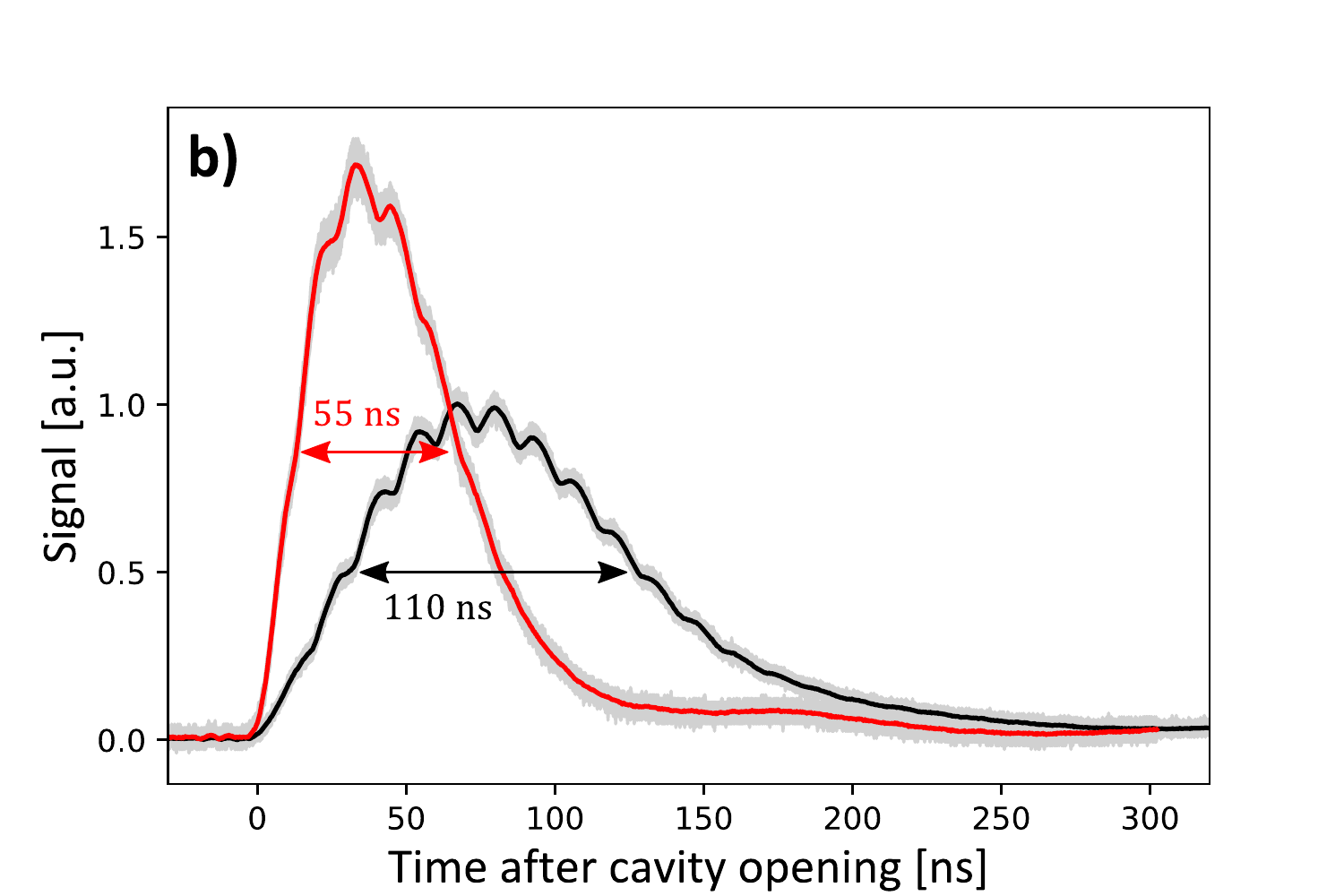}
\end{subfigure}
\caption{\textbf{a)} Effective reflectivity $R_\text{out}$ of the thin-disk laser resonator versus voltage difference $\Delta U$ across the BBO crystal in the Pockels cell. The dashed arrows indicate a switching sequence where the resonator is closed by switching the first electrode to $U_1=8$~kV and then opened by switching the second electrode to either $U_2=8$~kV (black, dashed) or $U_2=13$~kV (red, dashed). \textbf{b)} Typical pulse shapes for the two switching sequences in a). The gray shaded area shows the envelope of 100 pulses and the solid lines their average. Both pulses have an energy of 25~mJ. \label{fig:Switching}}
\end{figure}

A typical switching sequence is illustrated in Fig.~\ref{fig:Switching}. Initially, $U_1=U_2=0$~V, and the $\lambda /4$-plate is rotated so that $R_\text{out}(t<0) \approx 50$\,\% to allow seed incoupling while keeping the losses high enough to prevent pulse build-up. The TDL is PDH-stabilized in this state, and the intra-resonator power amounts to ca. 0.5~W. At $t=0$ the HV-switch to electrode E1 is closed so that $U_1 = 8$~kV builds up in about 10~ns. This results in $\Delta U = 8$~kV and the resonator is closed with $R_\text{out} \approx 100$\,\%. The pulse builds up quickly from the axial mode populated by the seed laser. After a pre-defined pulse build-up time, electrode E2 is switched to $U_2 = 13$~kV, resulting in $\Delta U = -5$~kV and an outcoupling reflectivity of $R_\text{out} \approx 25$\,\%. After 2~\textmu s the switches to E1 and E2 are opened and the electrodes discharge with a decay time $\tau \approx 500$~ns. Both electrodes are grounded so that $R_\text{out} \approx 50$\,\% is re-established. The TDL resonator length is stabilized again by the PDH lock and the laser is ready for the next trigger.

By controlling $U_1$ and $U_2$ independently, a lower outcoupling reflectivity can be achieved compared to the case when both electrodes are switched with the same voltage. This results in pulses with shorter tails which are more suited for subsequent non-linear conversion. As shown in Fig.~\ref{fig:Switching}b the temporal pulse shape significantly varied with the opening voltage $U_{2}$. For lower values of $U_{2}$ (e.g. $U_{1}=U_{2}=8$~kV) the reflectivity of the resonator is higher during the outcoupling process, which leads to pulses with a longer tail. We were able to tune the FWHM pulse length between 55~ns and 110~ns by tuning $U_2$ from 13.5~kV to 8~kV. We limited $U_{2}$ to 13.5~kV to avoid damaging the HV switches.

\subsection{Autoalignment and pulse energy stabilization \label{subsec:Stabilization}}
During operation slow drifts (over several hours) might misalign the TDL resonator. This not only reduces the pulse energy and beam quality, but also limits the pulse-to-pulse stability to around 2\,\% (RMS). We therefore installed an auto-alignment system controlling the horizontal and vertical tilt of the mirror "cav" in Fig.~\ref{fig:OsciSketch} via piezo-electric actuators. The error signal was obtained by monitoring the leakage light of the resonator-enhanced seed prior to a pulse through mirror $\text{M}_\text{fold}$ on a quadrant photodiode ($\text{PD}_\text{A}$) to stabilize the position of the laser beam at approximately the location of the disk.

The output pulse energy was further stabilized by monitoring the intra-resonator power during pulse build-up with a fast photodiode ($\text{PD}_{\text{E}}$ in Fig.~\ref{fig:OsciSketch}). Once a pre-defined voltage level $V_{\textnormal{max}}$ was reached, the PC was triggered to  extract a pulse. To prevent laser-induced damage due to trigger failure, we set a maximal pulse build-up time after which the PC is switched regardless of the photodiode signal. The outcoupling time jitter introduced by this stabilization method is marginal since the pulse build-up time only changes logarithmically with the pulse energy (i.e. the intra-resonator signal).

\section{Demonstration of injection-seeding and chirp measurement \label{sec:SingleFreq}}
\subsection{Heterodyne measurement setup}
We checked proper injection-seeding by performing heterodyne measurements from which the bandwidth of the TDL laser pulses could be determined, as well as their instantaneous frequency relative to the seed laser frequency. Following the scheme from \cite{white_control_2004} we extracted the relative frequency shift $\Delta\nu(t)$ between TDL laser pulses and seed via quadrature demodulation of a heterodyne signal.

\begin{figure}
\begin{centering}
\includegraphics[width=0.85\textwidth]{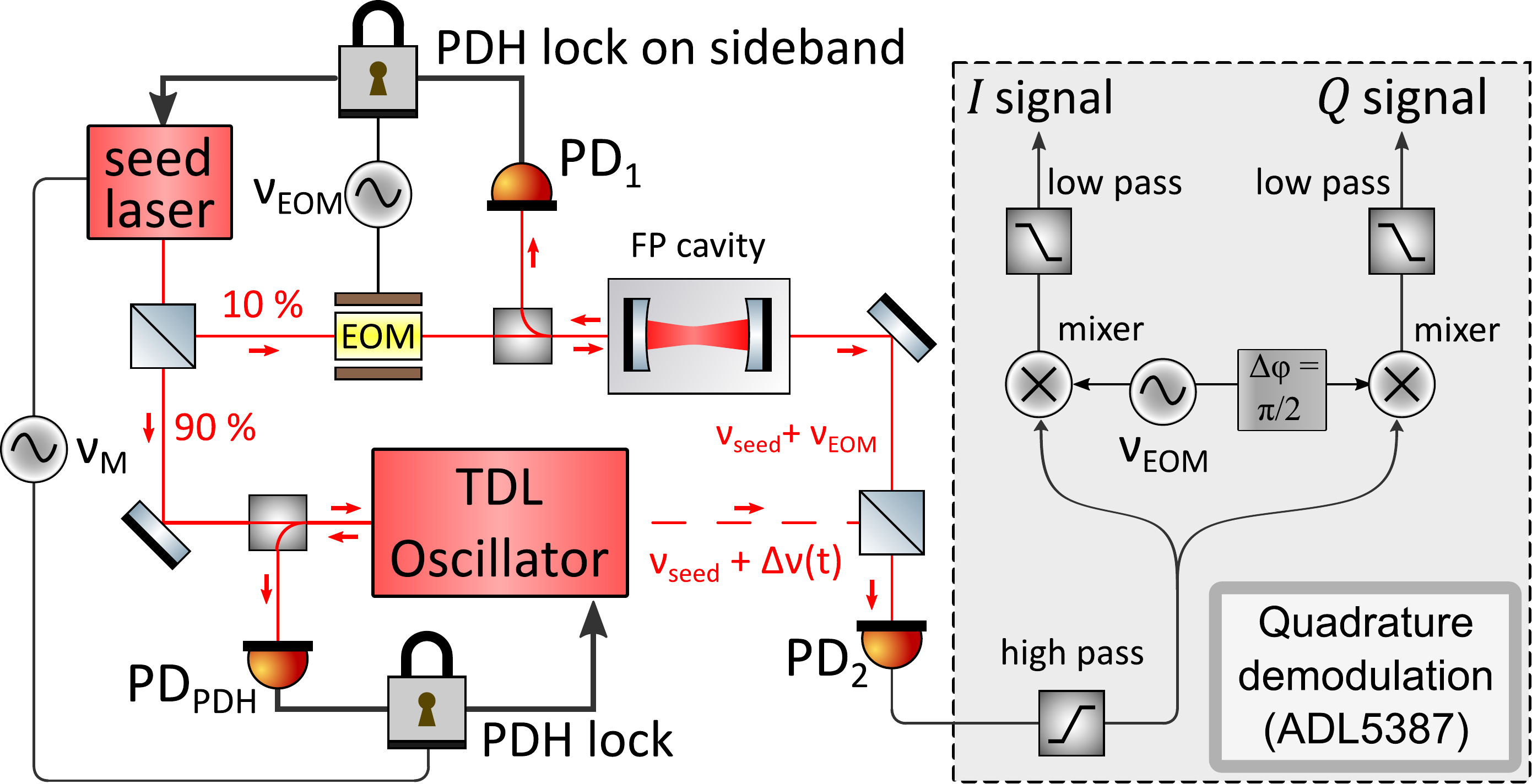}
\par\end{centering}
\centering{}
\caption{Sketch of the optical setup used to generated the heterodyne signal. The dashed red line represents the laser pulses. The seed laser output is split into two branches. One branch injection-seeds the thin-disk laser oscillator with photons at a frequency $\nu_{\text{seed}}$. The other branch contains an EOM imprinting sidebands at $\nu_\text{seed}\pm \nu_\text{EOM}$. The blue side band is kept resonant with the Fabry-Pérot (FP) cavity, so that only the frequency $\nu_\text{seed}+\nu_\text{EOM}$ is transmitted through the FP cavity and combined with the oscillator pulses on a fast photodiode ($\text{PD}_2$). The detected beat note is quadrature demodulated, and the resulting $I(t)$ and $Q(t)$ signals are used to measure the spectral properties of the TDL pulses. \label{fig:Chirp}}
\end{figure}
For the sake of brevity, we only sketch the experimental setup (see Fig.~\ref{fig:Chirp}) and present the result (for details see \cite{langenbach_realization_2021,zeyenthin}). The seed laser is current-modulated at $\nu_\text{M}=40$~MHz to lock the TDL to the seed laser with our modified PDH scheme. A fraction of the seed (10\,\%) is additionally phase modulated at $\nu_\text{EOM}=0.87$~GHz by a fiber-coupled EOM. This part is used to stabilize the seed laser frequency via a standard PDH lock on a Fabry-Pérot (FP) cavity so that the blue sideband (at $\nu_\text{blue}=\nu_\text{seed}+\nu_\text{EOM}$) is always resonantly transmitted through the FP cavity. The transmitted blue sideband is superposed with the pulse from the TDL on a fast photodiode ($\text{PD}_2$ in Fig.~\ref{fig:Chirp}). The beat note signal between the two beams recorded on this photodiode is high-pass filtered and sent to a quadrature demodulator board~(ADL5387).

\subsection{Pulse bandwidth extraction}
In this heterodyne measurement, the laser pulses had an exponential-like shape with a relatively long tail since the PC electronics were not yet modified to yield Gaussian-like pulses. The FWHM pulse length was $\tau = 35$~ns. Fourier transforming the square root of the high-pass filtered beat note signal and fitting its power spectral density around $\nu_\text{EOM}$ with a Lorentzian lineshape lead to a FWHM of $\Delta f = 3.7$~MHz as shown in Fig.~\ref{fig:Linewidth}a. This is in good agreement with the expected 3.1~MHz obtained from the time bandwidth product of $\tau \Delta f = \ln (2)/2\pi \approx 0.11$ for exponential pulses. The injection-seeded pulses from our TDL are thus close to Fourier transform limited. Although we have not repeated the measurement with the Gaussian pulses yet, we expect transform limited pulses in this case too.

\subsection{Chirp measurement}
\begin{figure}
\centering
\begin{subfigure}[b]{0.49\textwidth}
\centering
\includegraphics[width=\textwidth]{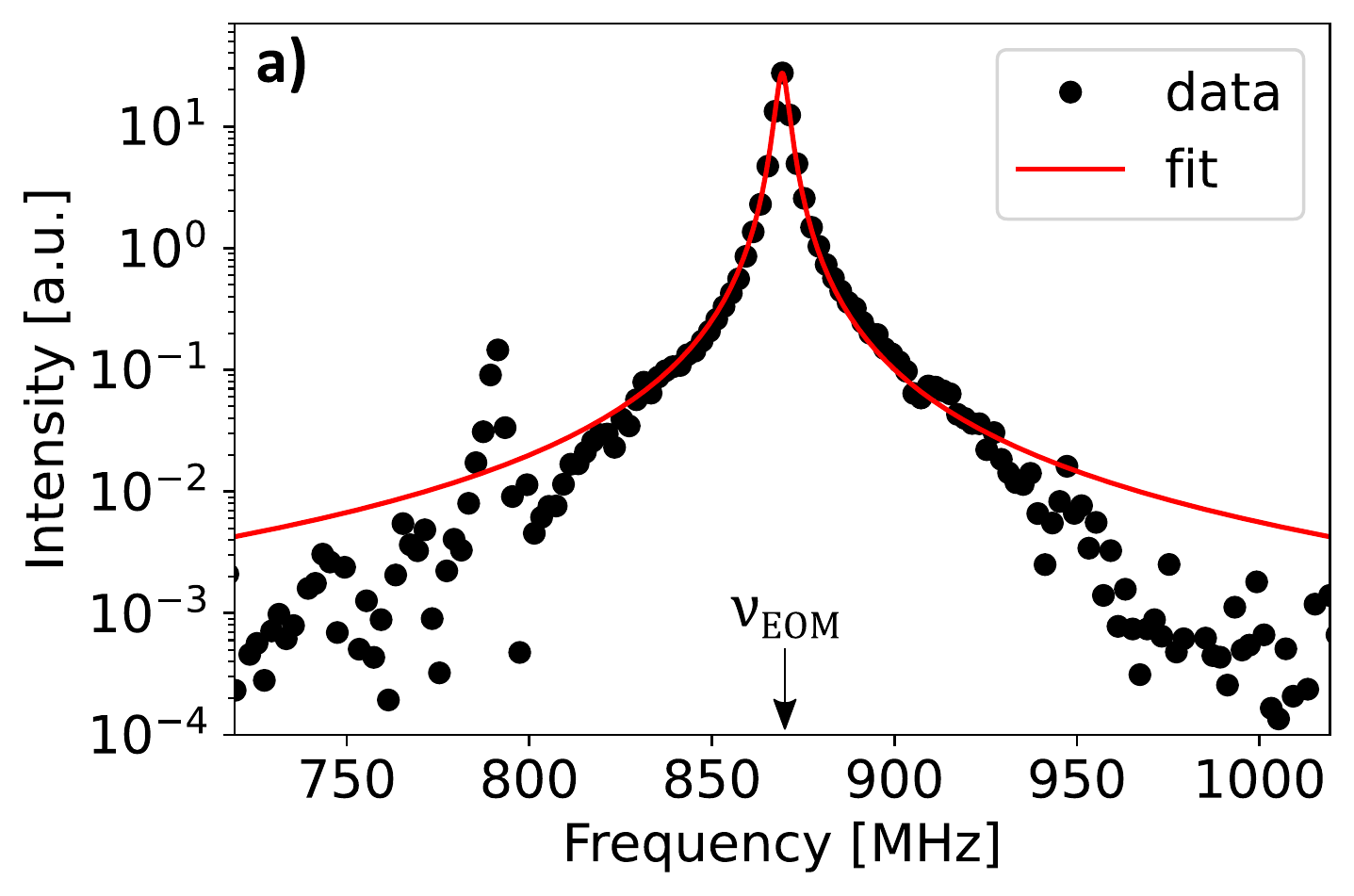}
\end{subfigure}
\hfill
\begin{subfigure}[b]{0.49\textwidth}
\centering
\includegraphics[width=\textwidth]{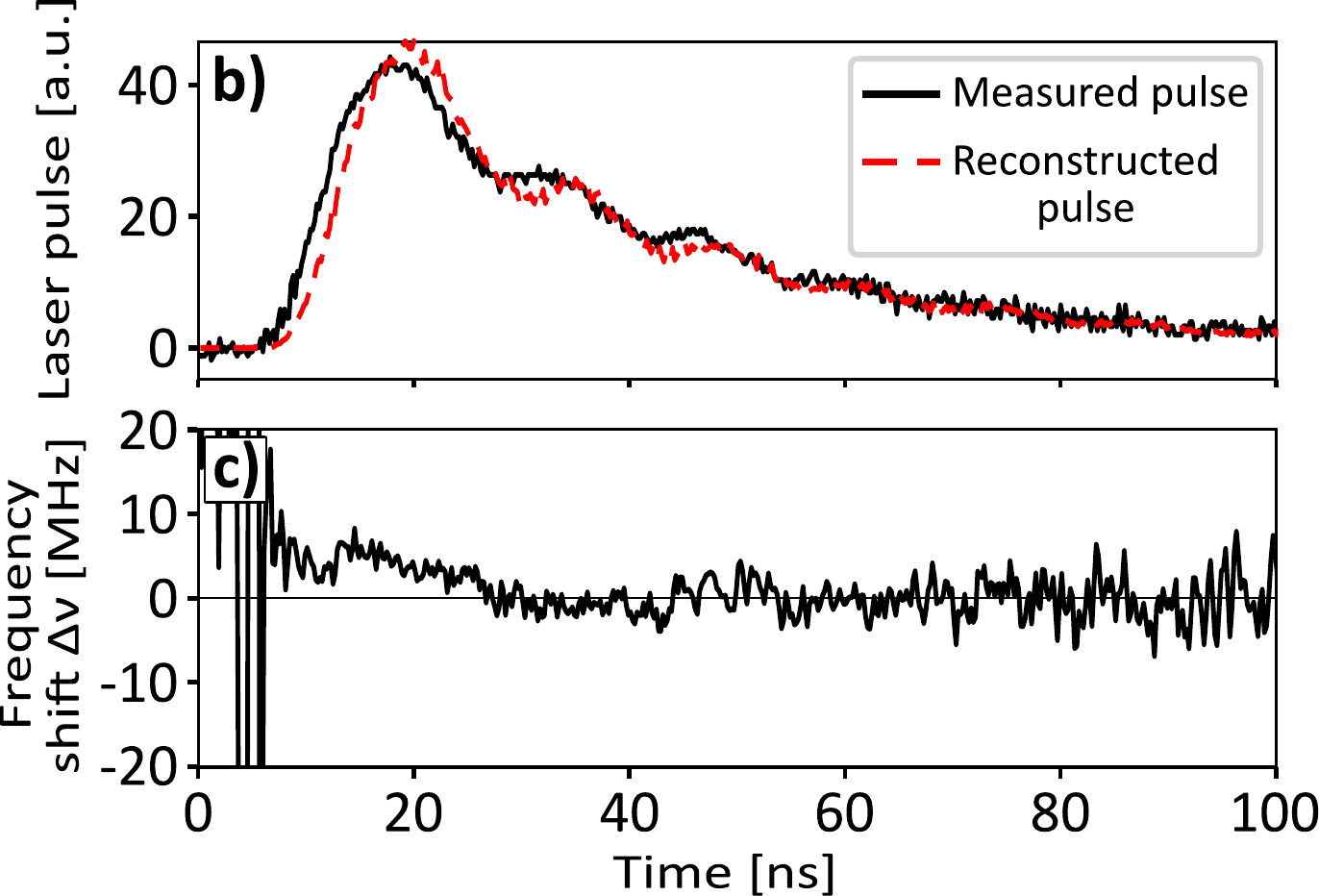}
\end{subfigure}
\caption{\textbf{a)} Power spectrum of the heterodyne beat note between the laser pulse and the frequency shifted seed laser for 10~mJ pulse energy. A Lorentzian profile with FWHM linewidth of 3.7~MHz was fit to the data. \textbf{b)} Top: Measured laser pulse (black, solid) and reconstructed laser pulse (red, dashed) using quadrature demodulation techniques for a pulse of 10~mJ energy. \textbf{c)} Instantaneous frequency shift between laser pulse and seed laser obtained from Eq.~\eqref{eq:inst_f}.
\label{fig:Linewidth}}
\end{figure}

The in-phase and quadrature signals $I(t)$ and $Q(t)$, obtained after quadrature demodulating the beat note signal, were used to reconstruct the pulse envelope via 
\begin{equation}
\left|A(t)\right|^2 \propto Q^2(t) + I^2(t)\label{eq:A(t)^2}
\end{equation}
and to calculate the instantaneous phase
\begin{equation}
\phi(t)=\arctan\left(\frac{Q(t)}{I(t)}\right).
\end{equation}
The latter allows extraction of the instantaneous frequency shift w.r.t. the seed light (i.e. the chirp of the pulse) through 
\begin{equation}
\Delta\nu(t)=\frac{1}{2\pi}\frac{d\phi(t)}{dt}.\label{eq:inst_f}
\end{equation}
Figure~\ref{fig:Linewidth}b shows a comparison between the original pulse (solid, black) and the reconstructed pulse (dashed, red) using Eq.~\eqref{eq:A(t)^2}. The close similarity of both pulse shapes serves as a cross-check of the single-frequency character of the original pulse since such a good reconstruction of the pulse can only be achieved if the original pulse is single-frequency. The bump-like features visible in the pulse are a general phenomenon which appears if the resonator losses are mostly localized at a single element (in our case the TFP) \cite{pollnau_spectral_2020}. The frequency of the resulting amplitude modulation lies inevitably at the TDL's free spectral range, and leads to a frequency component in the power spectrum at $\nu_\text{EOM} \pm 80$~MHz. A possible beating between two adjacent axial modes is thus masked by the bump-like features in the pulse. Nevertheless, as can be seen in Fig.~\ref{fig:Linewidth}a the spectral purity of the injection seeded TDL is better than 99\,\%. As shown in Fig.~\ref{fig:Linewidth}c, the frequency of the pulse was within $\pm 10$~MHz of the seed frequency at all times during the pulse. The increase of $\Delta \nu(t)$ towards the end of the pulse is due to noise since both~$I$~and~$Q$ signals are small. The wildly oscillating signal before the pulse is either related to the switching of the PC or a measurement artefact and requires more investigation. However, since it happens before the high intensity part of the pulse, it is not of practical relevance to the spectroscopy experiment.

\section{Laser performance\label{sec:LaserPerformance}}

Having demonstrated injection-seeding of the TDL, we characterized the laser performance in view of our muonic hydrogen spectroscopy experiment.


\begin{figure}
\centering
\includegraphics[width=\textwidth]{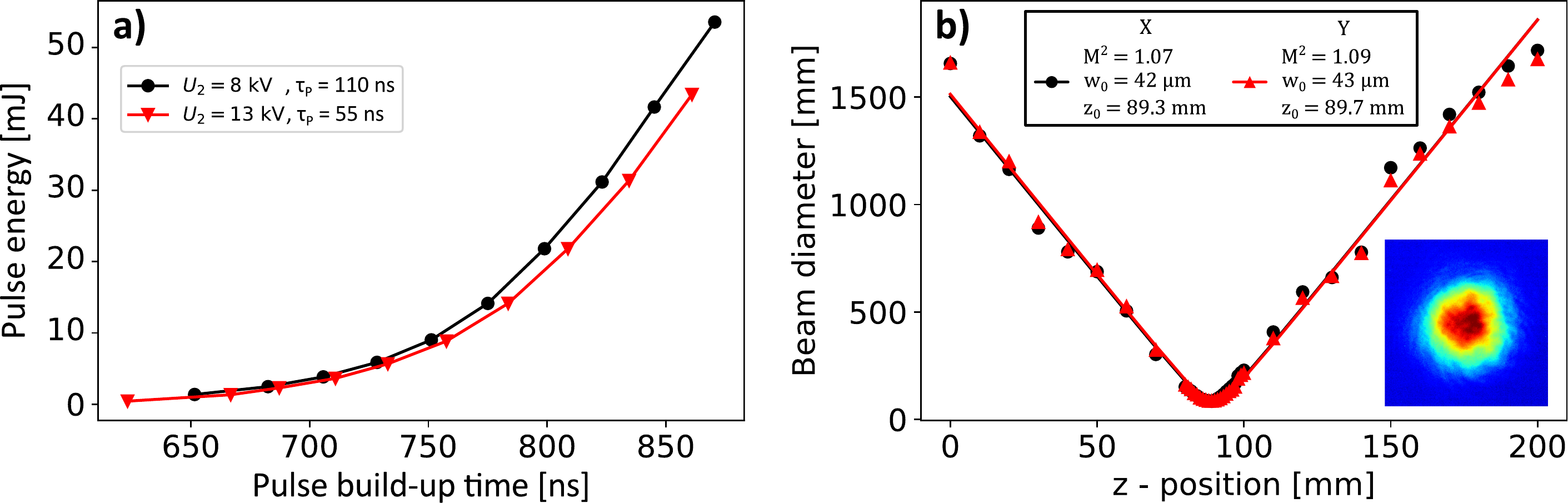}
\caption{\textbf{a)} Pulse energy delivered by the injection-seeded thin-disk laser versus pulse build-up time for two opening voltages $U_2$ at the Pockels cell leading to different FWHM pulse lengths $\tau_\text{P}$. Each data point represents the average over 500 pulses.  The pump spot has a diameter of $d_\text{P} = 5.8$~mm, the initial resonator reflectivity is $R_\text{out}=50$\,\% and the repetition rate is 30~Hz. \textbf{b)} $\text{M}^2$ measurement of 35~mJ pulses delivered by the injection-seeded thin-disk laser at 30~Hz repetition rate. The beam diameter was obtained by calculating the second-order intensity moments of the beam from pictures taken by a movable camera.
Inset: Near-field beam profile of a laser pulse with 1/$\text{e}^2$ radius of 2.1~mm. \label{fig:Energy_M2}}
\end{figure}

Figure~\ref{fig:Energy_M2}a shows the exponential increase of the pulse energy versus pulse build-up time after trigger. We limited the pulse energy to around 50~mJ in order to avoid laser-induced damage of the TDL. The pulse energy is slightly higher at longer pulse duration (black, circular markers) since the tail of a longer pulse is also amplified during the outcoupling. Thanks to the cavity layout and the injection-seeding, the beam is close to diffraction limited with $M^2 \leq 1.1$ (Fig.~\ref{fig:Energy_M2}b).
\begin{figure}
\centering
\includegraphics[width=\textwidth]{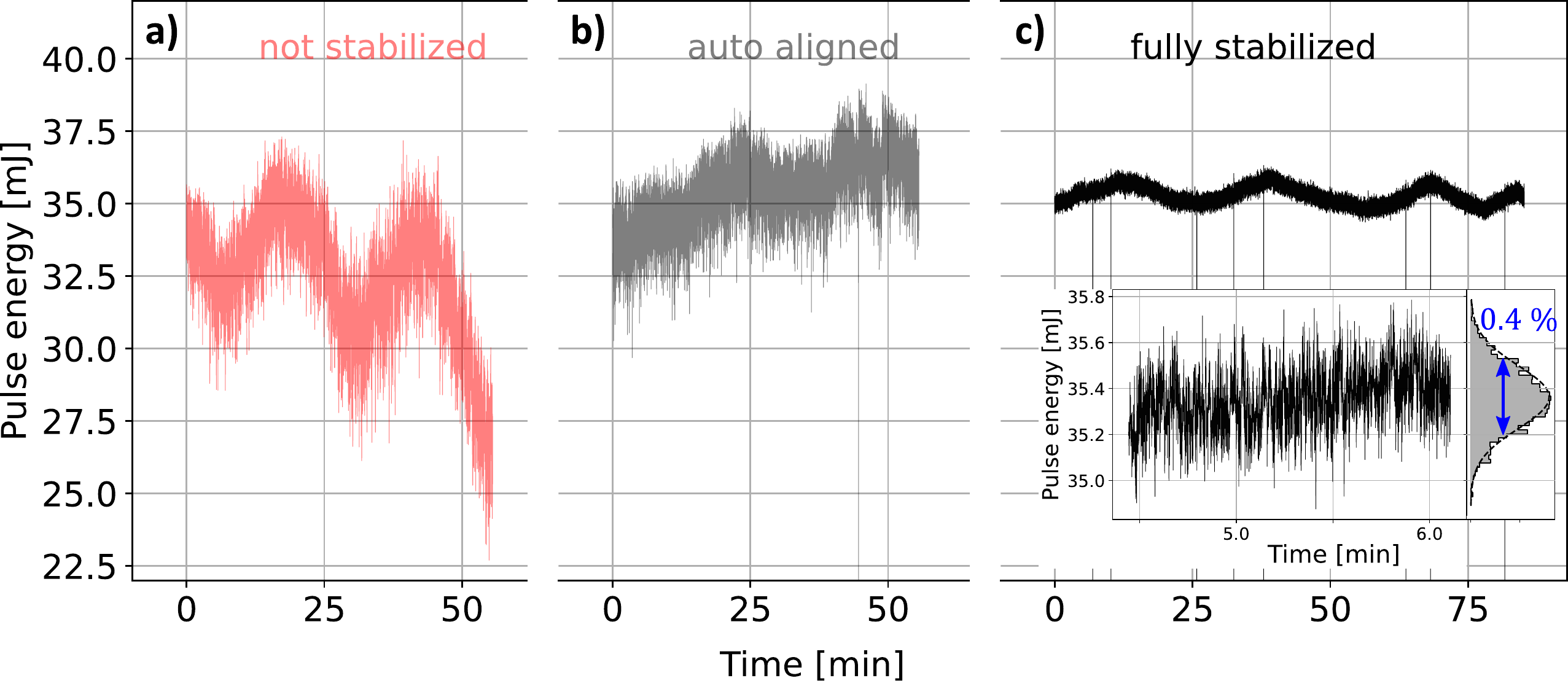}
\caption{Long-term measurement of the output pulse energy of the injection-seeded thin-disk oscillator \textbf{a)} without stabilization, \textbf{b)} with auto-alignment on, \textbf{c)} with both auto-alignment and pulse energy stabilizer on. With full stabilization the peak-to-peak variation is 4\,\% over about 80~min of operation. The inset is a zoom on 2 minutes of operation in the fully stabilized case. On this time scale the pulse-to-pulse energy variation is around $0.4$\,\%. The sudden drops in pulse energy visible in \textbf{b)} and \textbf{c)} occur when the laser re-locks.}
\label{fig:Stability}
\end{figure}
As shown in Fig.~\ref{fig:Stability}, the pulse-to-pulse energy variation is below 2\,\% (RMS) without additional stabilization, but the pulse energy exhibits drifts on the order of 20\,\% over time scales of hours due to seed power fluctuations related to variations of the polarization in the fiber and mechanical drifts caused by air temperature and cooling water temperature fluctuations. The auto-alignment is able to suppress these longterm fluctuations and even allows us to operate the TDL immediately from a cold start. It also slightly reduces the short-term pulse-to-pulse energy variation. The pulse energy stabilization further improves the pulse-to-pulse energy stability to <\,$0.5$\,\% (RMS) while increasing the timing jitter only marginally (<\,$0.3$~ns), so that the TDL can run reliably for hours. The remaining energy fluctuations on the order of $<4$\,\% (peak-to-peak) are most probably due to temperature instabilities in the cooling water and the air conditioning and require further investigation. The sudden drops in pulse energy visible in Fig.~\ref{fig:Stability}b and c correspond to re-locking of the injection-seeded TDL. Manifestly, a re-lock does not affect the stability of the laser.
\begin{figure}
\centering
\begin{subfigure}[b]{0.49\textwidth}
\centering
\includegraphics[width=\textwidth]{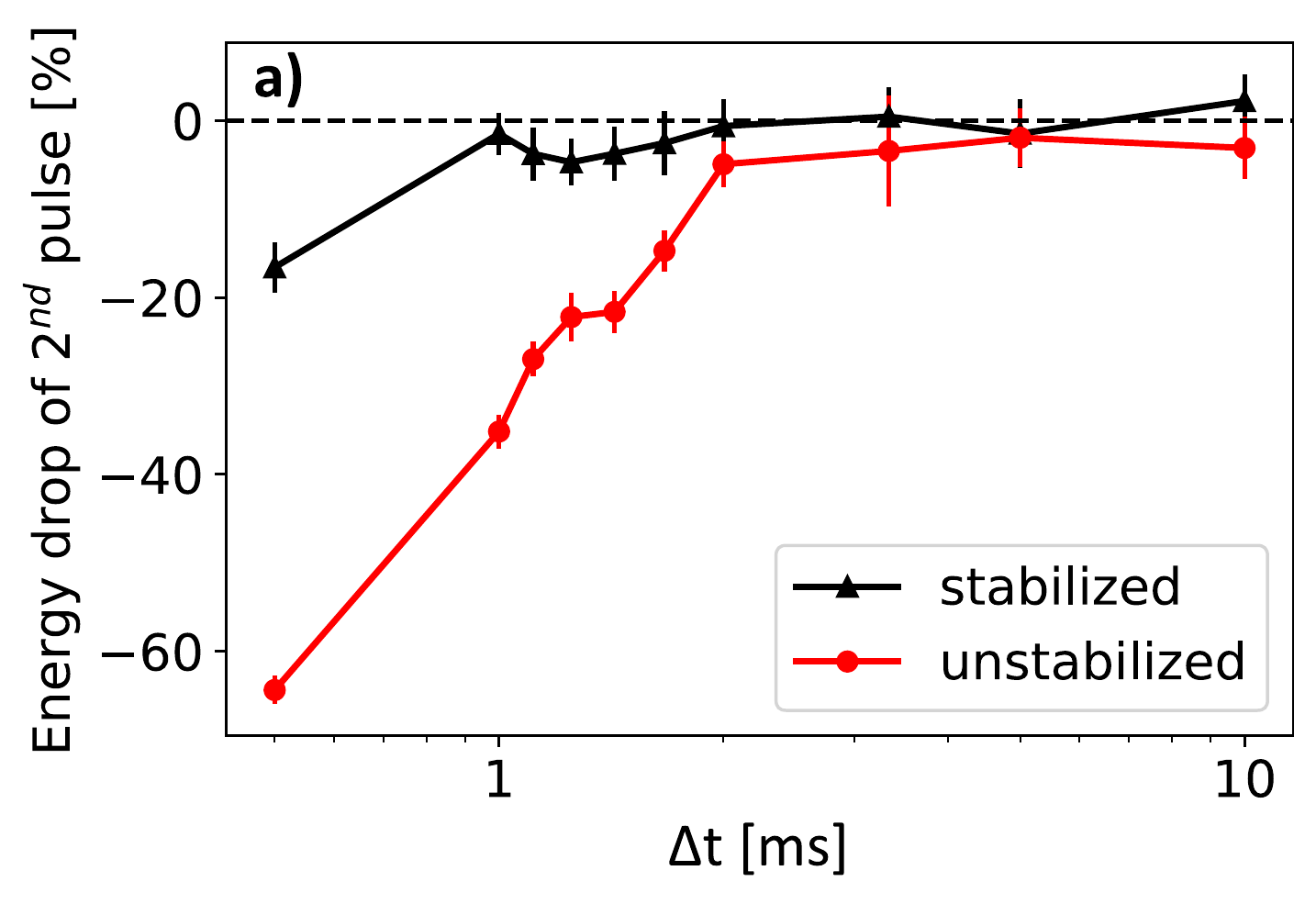}
\end{subfigure}
\hfill
\begin{subfigure}[b]{0.49\textwidth}
\centering
\includegraphics[width=\textwidth]{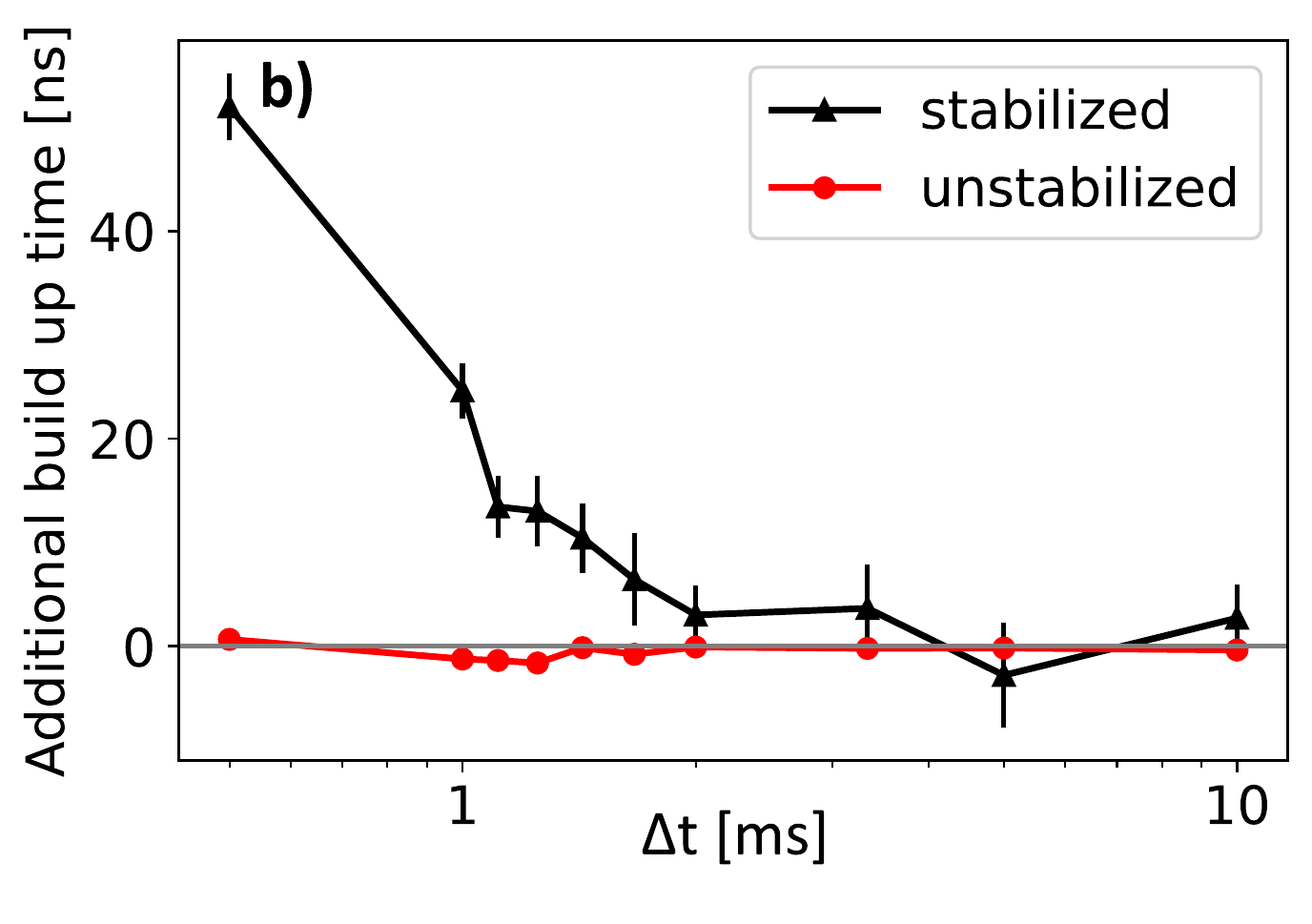}
\end{subfigure}
\caption{\textbf{a)} Energy drop of the second pulse relative to the first pulse $\left(E_2-E_1\right)/E_1$ versus time $\Delta t$ between the two pulses with (black, triangles) and without (red, circles) pulse energy stabilizer. The error bars are obtained from about 300 double-pulses recorded for each setting of $\Delta t$. In this data series $E_1 = 14$~mJ and the overall repetition rate was 30~Hz.
\textbf{b)} Additional pulse build-up time introduced by the pulse energy stabilizer for the measurements given in panel \textbf{a)}. \label{fig:DoublePulse}}
\end{figure}

For the muon spectroscopy experiment the minimal time between two pulses without significant reduction of the pulse energy is important. We thus tested the laser in double pulsed mode with variable time $\Delta t$ between two pulses at an overall repetition rate of $f_\text{rep} = 30$~Hz. The drop in pulse energy from the first pulse with energy $E_1$ to the second pulse with energy $E_2$ is shown in Fig.~\ref{fig:DoublePulse}a as a function of the time $\Delta t$ between first and second pulse. With these operating conditions, the pulse energy stays roughly constant down to $\Delta t = 2$~ms, which corresponds to a maximal triggering rate of about 500~$\text{s}^{-1}$. For smaller $\Delta t$, which lead to smaller population inversion for the second pulse, the pulse energy stabilizer compensates the decreased gain by increasing the pulse build-up time as illustrated in Figure \ref{fig:DoublePulse}b. For example, at $\Delta t=1$~ms the added pulse build-up time is 24~ns. The drop in pulse energy can be avoided down to $\Delta t = 1$~ms if the pulse energy stabilization unit is switched on (black, triangular data points in Fig.~\ref{fig:DoublePulse}a. Hence, the pulse energy stabilizer can be used to significantly increase the average repetition rate of the laser system when stochastically triggered.

The drop in pulse energy at $\Delta t = 0.5$~ms even with the energy stabilizer turned on is explained as follows: The energy stabilizer keeps the resonator closed for some time $t_\text{stab}$ until a certain value of intra-resonator intensity is detected. In reality, however, the resonator is only opened after a time $t_\text{stab} + \delta t$ where $\delta t$ is an additional fixed delay caused by the stabilizer electronics. For very short pulse separations $\Delta t$ the gain for the second pulse falls so much behind the gain of the fist pulse that the energy extracted in this additional time $\delta t$ is much smaller than for the first pulse.

\section{Conclusion and outlook\label{sec:Discussion}}

We developed an injection-seeded, Q-switched, Yb:YAG thin-disk laser with the following properties:
\begin{itemize}
\item $\text{TEM}_{00}$ output with $\text{M}^2 \leq 1.1$,
\item Up to 50~mJ pulse energy (we limited the output energy to this value to avoid laser induced damage),
\item Stable output pulse energy, independent of the delay between pulses up to 1~kHz (at 14~mJ),
\item On demand pulse generation with latency of 850~ns from trigger to pulse delivery,
\item Adjustable pulse length from 55~ns to 110~ns,
\item Fourier limited pulses with $<5$~MHz linewidth,
\item Pulse-to-pulse energy fluctuations of 0.4\,\% and peak-to-peak energy fluctuation $<4$\,\% over hours of operation.
\end{itemize}

Single axial mode operation was achieved by PDH-locking the thin-disk laser resonator to a master seed laser. By modulating the seed laser at half the free spectral range of the thin-disk laser resonator, the PDH-lock was guaranteed to re-lock correctly even after large disturbances. In the present setup, the pulse energy can be stabilized up to a repetition rate of about~1~kHz which is well beyond our required repetition rate. If needed, the pulse build-up time could be reduced by shortening the thin-disk laser resonator length, by increasing the pump power density (i.e. the gain in the disk) and by increasing the seed power. We plan to further increase the long-term stability of our system by improving its temperature stability.
This single-frequency thin-disk laser will be used in the laser system for the measurement of the ground-state hyperfine-splitting in muonic hydrogen, currently being pursued at the Paul Scherrer Institute in Villigen, Switzerland.

\section*{Funding}

The European Research Council (ERC) through CoG. \#725039; the Swiss National Science Foundation through the projects SNF 200021\_165854 and SNF 200020\_197052; the Deutsche Forschungsgemeinschaft (DFG, German Research Foundation) under Germany's Excellence Initiative EXC 1098 PRISMA (194673446), Excellence Strategy EXC PRISMA+ (390831469) and DFG/ANR Project LASIMUS (DFG Grant Agreement 407008443); the French National Research Agency with project ANR-18-CE92-0030-02;

\section*{Acknowledgments}
We gratefully acknowledge the support of the ETH Zürich electronics workshop. In particular, we thank Diogo Di Calafiori and Dr. Werner Lustermann. We also thank Dr. Marcos Gaspar, Dr. Carlo Vicario, Dr. Cezary Sidle and Stefan Mair from PSI.

\section*{Disclosures}

The authors declare that there are no conflicts of interest related to this article.

\section*{Data Availability Statement}

The data that support the findings of this study are available
from the corresponding author upon reasonable request.

%

\end{document}